# What If People Learn Requirements Over Time? A Rough Introduction to Requirements Economics

**Corentin Burnay**
Namur Digital Institute, University of Namur, Belgium
corentin.burnay@unamur.be

**Ivan Jureta**
Fonds de la Recherche Scientifique - FNRS,
and Namur Digital Institute, University of Namur, Belgium
ivan.jureta@unamur.be

## Abstract

The overall objective of Requirements Engineering is to specify, in a systematic way, a system that satisfies the expectations of its stakeholders. Despite tremendous effort in the field, recent studies demonstrate this is objective is not always achieved. In this paper, we discuss one particularly challenging factor to Requirements Engineering projects, namely the change of requirements. We proposes a rough discussion of how learning and time explain requirements changes, how it can be introduced as a key variable in the formulation of the Requirements Engineering Problem, and how this induces costs for a requirements engineering project. This leads to a new discipline of requirements economics.

## 1. Introduction

Requirements Engineering (RE) designates the practice of, and research on acquisition, generation, analysis, specification, and validation of requirements for a system-to-be.

The term “engineering” suggests that the process should be systematic. Research output is indeed often normative, suggesting how to do any of many activities one must do when doing RE. The overall aim is to produce such specifications, that the resulting system, once up and running, satisfies the expectations of its stakeholders. That is has both high engineering quality, i.e., runs well according to specification, and high service quality, by meeting and perhaps exceeding its users' expectations.

This is far from being the case in practice. Even today, more than 70 percent of information system projects are being somehow challenged, meaning they do not reach quality or resource use targets (SG2016). Among the top contributing factors, change of requirements and specifications is a major one.

Failing to account for changes increases the risk of failure of the project, according to the Standish Group. This is not a new idea; the same risk was noted by Boehm, in a 1979 paper (B1979); "*IBM's Santa Teresa software organization has found, on a sample of roughly 1'000'000 instructions of software produced per year to IBM-determined requirements, that the average project experiences a 25% change in requirements during the period of its development*".

It seems obvious to observe that requirements can, and often do change over time. But identifying the consequences of that idea on the very conceptualization of RE - of its core concepts and methods - is far from trivial.

Isn't that already accounted for in RE research? It is. Work on requirements evolution (EBJ2011,EBJM2012,JBEM2015) focuses predominantly on how to anticipate change or evolution requirements, and how to adapt a specification accordingly, and in some limited ways propagate those changes to the system configuration at runtime.

In work on requirements evolution, time is an important, but secondary variable. What matters are changes, and specifically changes of requirements, since they are what triggers adaptation, i.e., specification change.

In this paper, we go back to basics, introduce learning of requirements and how it is influenced by time, and discuss consequences from one specific perspective more than others. That perspective is cost that change induces. As we hope to show in the rest of the paper, this leads to a new discipline of requirements economics.

By basics, we mean the basic requirements problem statement. Zave and Jackson provided a synthetic formulation of this problem, which is at the same time simple enough for us to develop our main argument without having to digress much.

The problem is stated as follows. Given some domain assumptions $K$ and some requirements R, the problem is to design a specification S such that K and S satisfy R (ZJ1997). With its split between K, S, and R, the requirements problem provides a simple ontology which defines the set of concepts that should be accounted for by any RE notation, model or more generally any RE theory. It has been and remains central to a considerable part of RE research.

# 2. Environment and Time

We start the discussion here with a classical idea in Requirements Engineering. Namely, there is something called the Environment. It denotes that part of the world that is relevant, and a given to stakeholders and requirements engineers (Jackson, 1997). Environment can be seen as the context in which the system-to-be will run, in order to provide some services to stakeholders. We denote the *actual state of the environment as* $E$. $E$ is actual in the sense that it is the true world state, so to speak, so not someone's perception of what is or is not true. It is also not a representation of the environment.

When E denotes past and present conditions, it is a given and cannot be changed. Taking the example of a mobile banking application, the environment $E$ would cover the bank itself as a business with a strategy and sales objectives, the smartphone, tablet and other hardware clients, the server, but could also include less tangible concepts such a law on financial transactions or any other financial regulation. At the same time, the environment would exclude such concerns as, for example, social network software which, even if it is used by the same people who use the banking application, is in no way directly related to that banking application, and, moreover, requirements engineers assume that there is no interaction between the usage of one and usage of the other (this might be a simplifying assumption, of course, but whether it is made, is up to the requirements engineers solving that specific problem).

The Environment changes over time. As time goes by, events happen, and the environment keeps changing. If we simplify, we go from some E1, to an E2, and so on.

Some of these changes may be under the influence of the actors mentioned so far, but many are not. As the future has not happened yet, and is thus by definition uncertain, any predicted future Environment need not obtain.

This can be written as follows. If $t_1$ and $t_2$ are two different points in time, then changes mean that $E(t_1) \neq E(t_2)$ , and more generally that $\forall (i \neq j) : E(t_i) \neq E(t_j)$.

In our example, changes may be due to changes in the banking organization, in the broader banking context, competitive conditions on the market relevant for the software being made, and so on.

We depict the following in Figure 1, where $\Delta E_j$ describes the changes occurring in the environment at time $j \leq n$, where $n$ represents the present. Note that, $f$ represents the moment when the system will be released to customers. Then, $E_f$ and any intermediate state of the environment between time $n$ and time $f$ is empty, i.e., we assume future states of the Environment cannot be known, so that $E_{j>n} = \emptyset$.

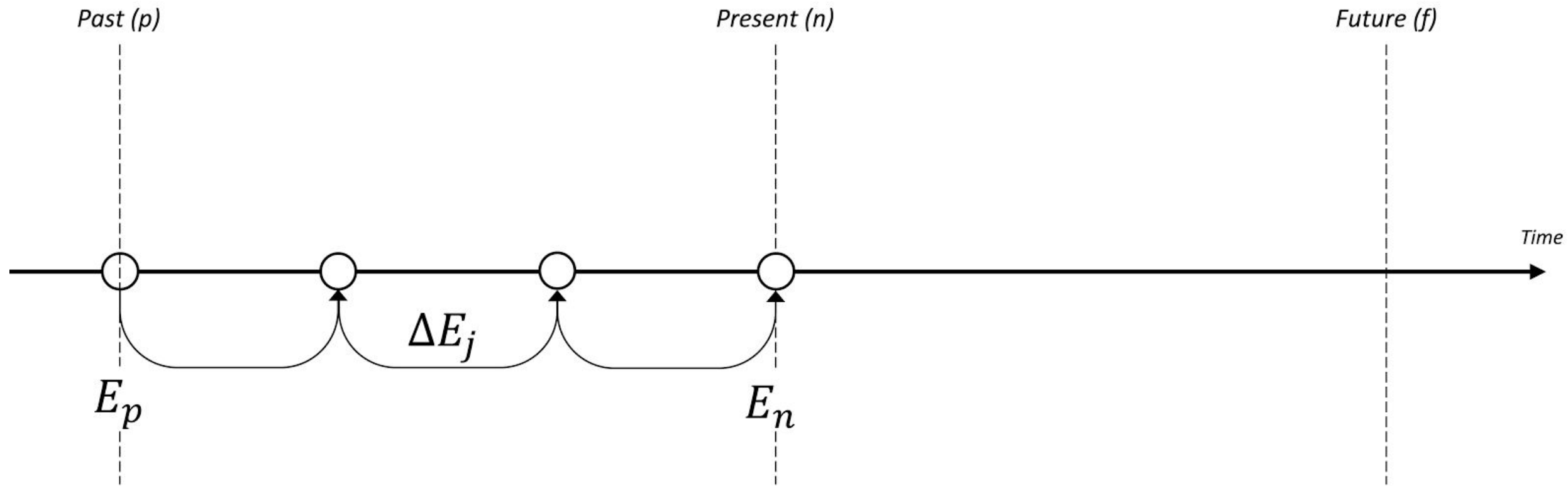

*Figure 1 - Environment, Changes and Time*

# 3. Attainability and Access

We now want to distinguish between the perfect information about the true Environment, and imperfect information about it. The former is a theoretical notion; current science does not explain all there is to explain, and an individual can only know so much. Simply stated, if there is some theoretical notion of perfect information about the Environment, then all we can access is some incomplete part of that. We will say that there is:

- Perfect information about the Environment, which would be all truths we could ever know about the Environment, if we could know that at all;
- Attainable information, which is that part of Perfect information, which could, provided unlimited resources, be found out; this is again too much, since we do not have unlimited resources[1];
- Accessed Attainable information, which is that part of Attainable information which we managed to actually find out, by investing resources to do so.

We represent this idea graphically with a Venn diagram in Figure 2. Attainable information $I_n^A$ is function of the state of the environment: $I_n^A = Att(E(n))$. Since $I_n^A$ is function of $E$ which is function of $n$, $I_n^A$ is indirectly function of time, i.e., $I^A(E(t_1)) \neq I^A(E(t_2))$.

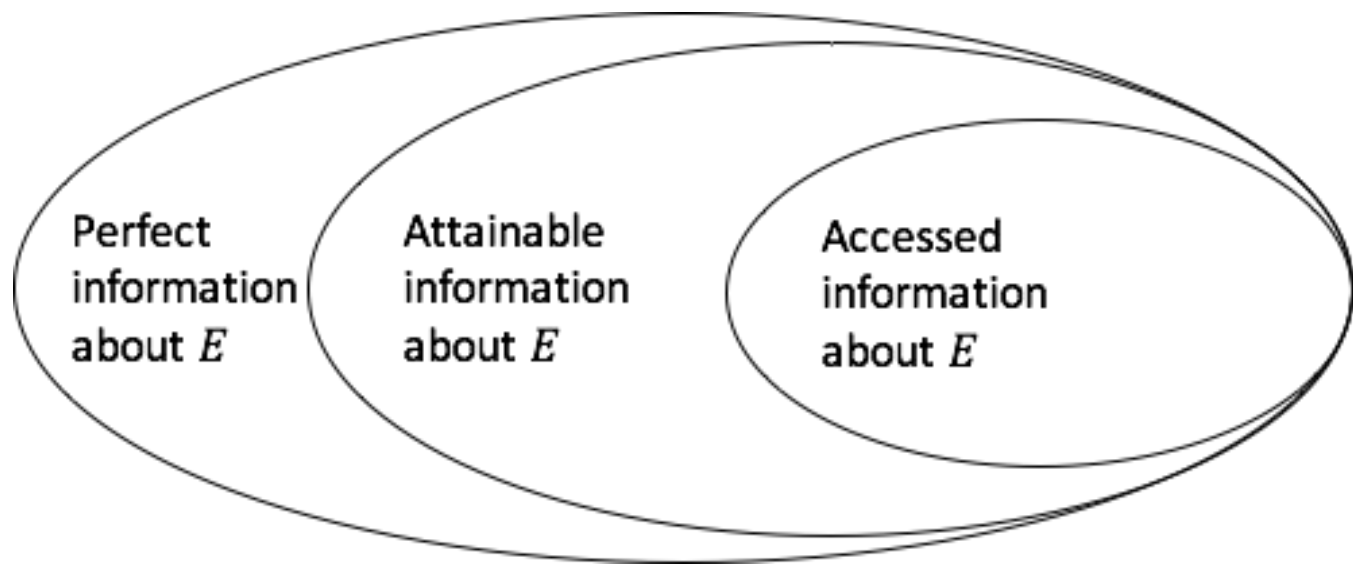

[1] We consider that Attainable information include only information that is true. This excludes false information believed by stakeholders (due, for instance, to lies or wrong conclusions).

*Figure 2 - Perfect, Attainable and Accessed Information*

*Attainable information is not fully accessed*. In many situations, part of the attainable information is not accessed and used by people when reasoning (we will come back to this reasoning later in the paper).

Unlike attainable information, accessed information depends on some human factors and time (the two being closely interrelated). While it is not limited to these, human factors include such things as memory and attention, both of which play a role in which information one would access, when more information is available.

Memory means that information in human memory (not necessarily computer memory) decays. Information about older states of the environment is more difficult to access. On the contrary, more recent states of the attainable environment are easier to access. This brings us to the definition of the accessed information $I_n^a$, which is function of the attainable information, i.e., $I_n^a = Acc(I_n^A(E(n)), Inv_\alpha)$. Note that we suggest the accessed information is also function of an investment $Inv_\alpha$, reflecting the amount of resource allocated to the process of accessing information. The following is represented graphically in Figure 3.

We return to the banking application. It is reasonable to assume that information about similar projects carried out by competitors cannot be attained, because hidden. Similarly, information about the subjective factors that make people reluctant to use a mobile application as a way to manage their bank accounts is not attainable because held in mind, so to speak.

These information are part of $E$, likely matter to Requirement Engineers but are not part of $I^A$, i.e., they are not part of attainable information.

Going further, a person may not be able to access all the attainable information. She may for instance remember that a similar project took place two month ago and failed. She can however omit a one year old discussion when it was decided that “making customers’ life easier” was the main goal of the project, thereby further reducing $I^a$.

This piece of information could still be accessed if the stakeholder was to consent to invest some resource (money or time for instance) in the process, and that investment would amount to discussion with her colleagues, checking old memos, or simply trying to remember and concentrating on what has been discussed in the past, i.e., by increasing $Inv_\alpha$.

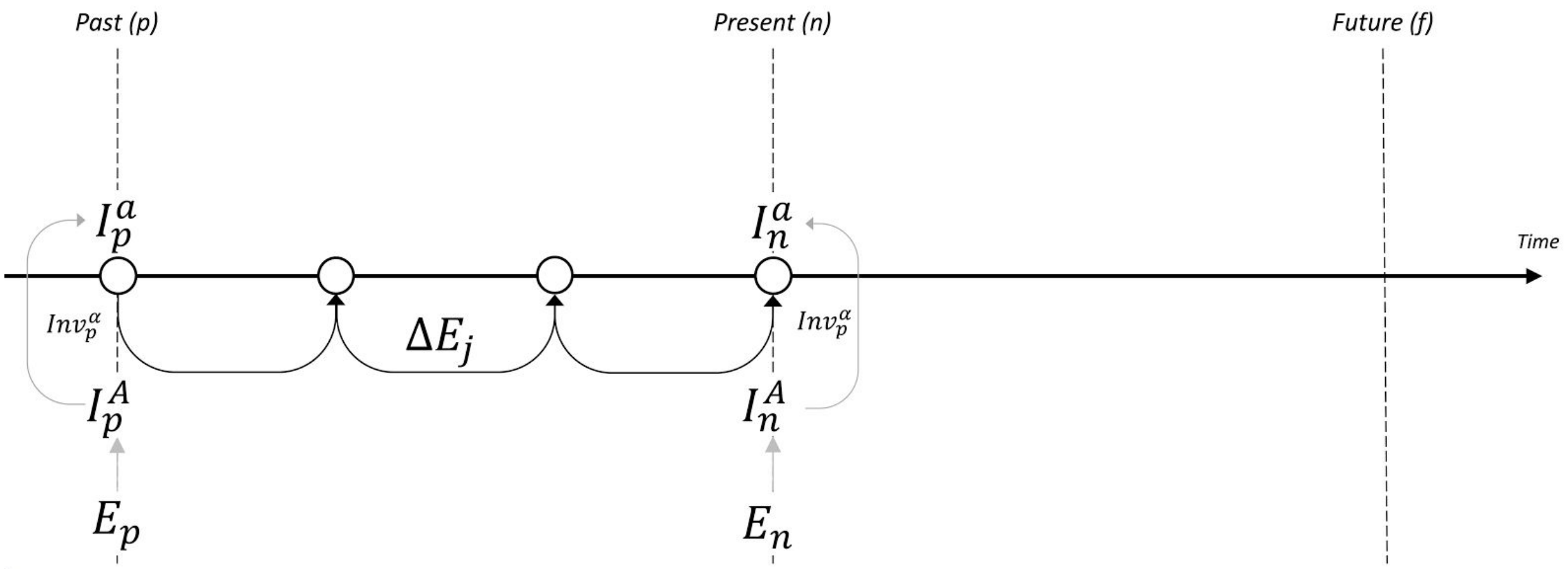

*Figure 3 - From Environment to Accessed Information*

# 4. Recorded and Unrecorded Attainable Information

An important parameter that influences the extent to which an attainable information can be accessed is the representation of information. In fact, we believe a clear distinction has to be made between the attainable information that is recorded (explicitly represented) and unrecorded (not explicitly represented). We discuss this below.

*Much of attainable Information is going to be unrecorded*. Unrecorded means that the information is not transcribed physically on a support, be it a technological (numerical) support like a database or a physical support like a sheet of paper. In other words, it is not documented, and as a consequence is less permanent. It does not mean that the information cannot be accessed. In practice, the information remains in the memory of (at least) one person. It means however that the risk is higher of seeing this information becoming unaccessed, due to memory decay.

*Attainable Information is sometimes recorded*, in a database, an email, a memo, a previous requirements analysis, a model, a knowledge base, etc. During the operational activities of a company, a piece of information may be recorded, because a procedure tells the worker to do so, because the worker judges this might be valuable someday, or simply because people discussed it via a technological medium (using a chat or a wiki for instance). The fact that an information is recorded makes it (more) permanent, and more likely to be easily accessed. However, it requires some investment by people in the company to actually write things down. It also does not guarantee that the information is true, that is, that it was in fact part of Attainable information in the first place; perhaps it is a wrong conclusion drawn from attainable information.

The former paragraphs suggests that $I^A(E(n))$ is actually made of unrecorded information and some recorded information - provided some effort has been consented to actually record it, i.e., $I^A(E(n)) = I^{A,Unrec}(E(n)) \cup I^{A,Rec}(E(n), Inv_\beta)$ , where $Inv_\beta$ refers to the amount of resource invested by an organization in the systematic transcription of attainable information. This distinction actually matters, because minimizing the ratio of unrecorded against recorded attainable information likely maximizes accessed information $I^a$. Next section clarifies why it is important to have the largest possible $I^a$ in a Requirements Engineering context.

# 5. Learning From Accessed Information

Requirements do not pop up with no reason in stakeholders' minds. They do not exist a priori, ready to be elicited by requirements engineers. Instead, we claim requirements are the result of a long-lasting learning process of the stakeholders, who apprehend their environment - or more specifically the part of the environment they can actually access - and adapt to it. You have requirements for your email client to a large extent because you used them in the past.

For the sake of brevity in this paper, we call on the Perceptual Learning theory for an explanation of how requirements may be learned by stakeholders[2]. Goldstone claims that *"Perceptual learning involves relatively long-lasting changes to an organism's perceptual system that improve its ability to respond to its environment and are caused by this environment".* Translating this definition RE, we could say that requirements are learned through the perception of the environment; perceptual requirements learning occurs to improve a stakeholder's ability to respond to its accessible environment, and to changes occurring in that environment.

Perceptual learning makes sense in RE; requirements are relatively permanent when they build on practice and experience, i.e., a requirement based on a solid experience will likely be more reliable than a purely theoretical requirements, built on assumptions of the stakeholder. Moreover, Goldstone definition of Perceptual Learning implicitly suggests that learning occur because of some intention of the organism (the stakeholder) to adapt to a changing environment. Without any intention to adapt, no learning is likely to happen. This echoes research in Requirements Engineering about Intentional States, and how they shape requirements.

Compiling those different elements, we can say that requirements are produced through some *Lr* - "Learning Requirements" - function, which takes two different parameters. The first one is the union of all accessed information about past states of the environment, i.e., the experience, or knowledge background of the stakeholder. The second one is the set of intentional states of the stakeholders at time $t$, which also influences the way learning happens:

$$R_n = Lr\left(\bigcup_{time=-\infty}^{n} I^a(time),\ IS(n)\right)$$

[2] Note that the choice of this theory rather than another one has little impact on the theoretical work conducted in this paper; it simply provides some foundations on which to build our discussion.

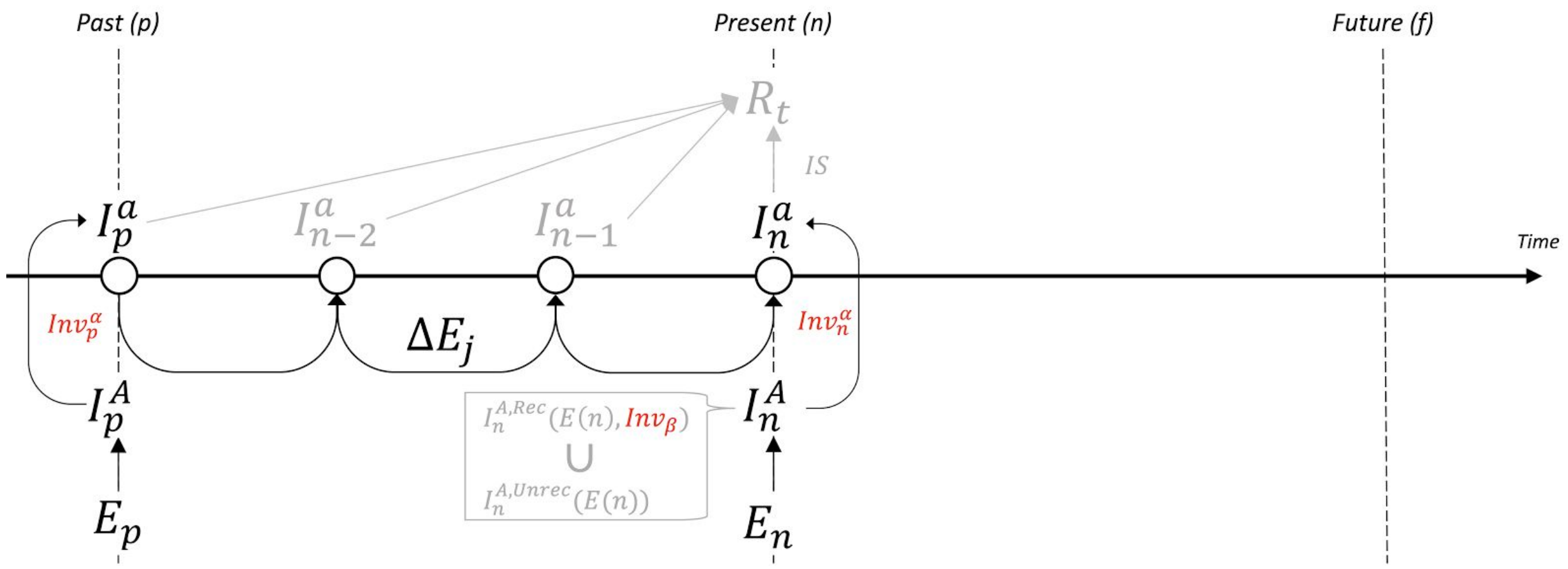

*Figure 4 - Learning Requirements from Accessed Information*

Going back to our running banking app example, it seems clear that a customer does not know, from scratch, what to expect from the app. She can either speculatively imagine (J2017) requirements or she may access present and past states of the environment and use them to define requirements. For instance, she has been repeatedly in a state where she wanted to purchase a present for a friend, but had no idea of the balance on her account. Without any certainty she could pay the gift, she delayed her purchase but missed a nice offer. Combined with a strong motivation to avoid this situation in the future (IS), she learns a requirement like “I must be able to check my account balance anywhere I want”. Figure 4 depicts this idea.

# 6. Investment and Maturity

When learning, it is difficult to achieve an acceptable result on the first attempt. The first time you try to do a complex manual task (e.g., pottery), you would likely come to realize it would take several rounds before achieving a quality result. Stakeholders are basically confronted to the same issue when learning requirements; learning is not a one shot process, that can be triggered at will by requirements engineers. It is a time consuming and often long lasting process, that demands concentration, application and maturity from stakeholders.

The main idea underlying previous paragraph is that the requirements produced at a certain time $t$ are likely influenced by previous (if any) requirements (for a given system). This brings us to a new proposition: the requirements learning function as defined in Section 4 has a third parameter, requirements maturity.

The more requirements a stakeholder shared in the past for a particular system, the more mature that stakeholder gets for that particular system’s set of requirements. Requirements maturity, however, cannot be defined solely based on the number of past similar requirements. For a stakeholder, asking for requirements without checking their relevance and usefulness will not contribute positively to requirements maturity. Instead, some effort should be made by a stakeholder to validate the requirements she shared; by validation, it is meant that the stakeholder tries to understand if her requirements actually helps her solve her problem, and if they make sense considering her attainable environment. Requirements maturity is therefore a function which takes as parameters all former requirements shared by the stakeholder for a given system, and the amount of resource invested in the validation of those past requirements (we denote it $Inv_\varepsilon$). We also define the *Lag of Maturity* as the delay between the moment a first requirement is shared for a given system and the moment the last requirement is shared for that same system. This idea is formalized in the equation below, and is also depicted in Figure 4.

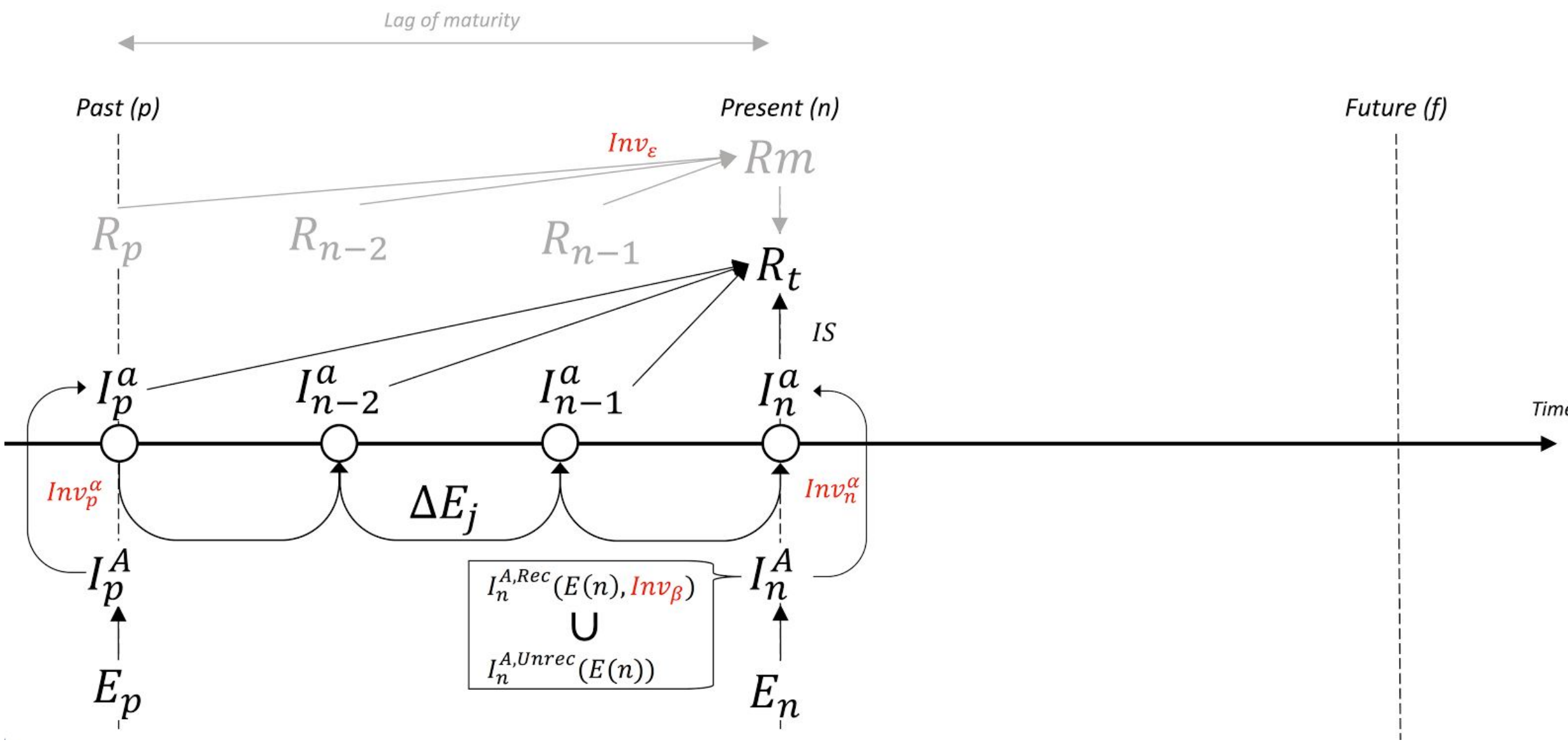

*Figure 4 - Learning Requirements from Requirements Maturity*

$$R_n = Lr\left[\bigcup_{time=n}^{-\infty} I^a(time),\ IS(n),\ Rm\left(\bigcup_{time=n-1}^{-\infty} R(time),\ Inv_\varepsilon\right)\right]$$

Let's illustrate this idea with our running example. Imagine the banking app user formulated in a previous elicitation round a requirement like "Once logged in, I must be able to make a payment without additional authentication". From there on, there are typically two possible scenarios. First, the stakeholder takes time to validate this requirements, to ensure it actually solves the problem at hand. Using a prototype, casting herself in various scenarios, she tries to find out if the requirement she shared is relevant or not. Second, she may simply state the requirement, without further attention. The impact on requirements maturity is likely to be very different in the first and second case. In the first case, she may realize that the requirement is in fact too vague, and that she wants to make un-authentified payments only to people who are registered as contacts in her app, for obvious security reasons she did not thought of when sharing her requirement at first. As a consequence, she gets more mature with regards to her requirements, and next (future) requirements will be impacted. In the second scenario, none of this happens, requirements maturity remains small, and future requirements remain the same, lower quality, requirements.

# 7. Learned Requirements Cannot Be Elicited Instantly

*Requirements learnt by stakeholders are elicited by requirements engineers at some cost*. Elicitation is an intensive communication process, which requires effort and investment from requirements engineers, who have to apply the broad range of elicitation techniques available to collect information. This is a well known issue in requirements engineering, and plethora of papers have been written about how to deal with it. That literature focuses on how to transfer information from stakeholders to engineers. This suggests a distinction has to be made between requirements learned by stakeholders ($R_n$, discussed in Section 4 and 5) and Elicited Requirement $R^e$. We claim is at best equivalent to learnt requirement, and more likely only a subset of it, $R^e \subseteq R_t$. Moreover, it takes resources to elicit requirements; the more resources are put in the elicitation process, the closer $R^e$ will get to $R_n$.

*Requirements learned by stakeholders cannot be transferred instantaneously to requirements engineers*. The elicitation process requires time. Most elicitation techniques, like interviews, workgroups, JAD sessions and other workshops are time consuming. This introduces a lag between the moment a requirement is learnt by a stakeholder and the moment that same requirement is documented by an engineer. Even high investment levels (e.g., assigning several engineers to the elicitation process) will not completely suppress this Lag of Elicitation. Taking considerations from previous paragraphs into account, we can say that $R^e_j = Elicit(R_n, Inv_\theta)$, where $j > t$ and reflects the Lag of Elicitation, and $Inv_\theta$ represents the level of investment in the elicitation process. This is represented graphically in Figure 6.

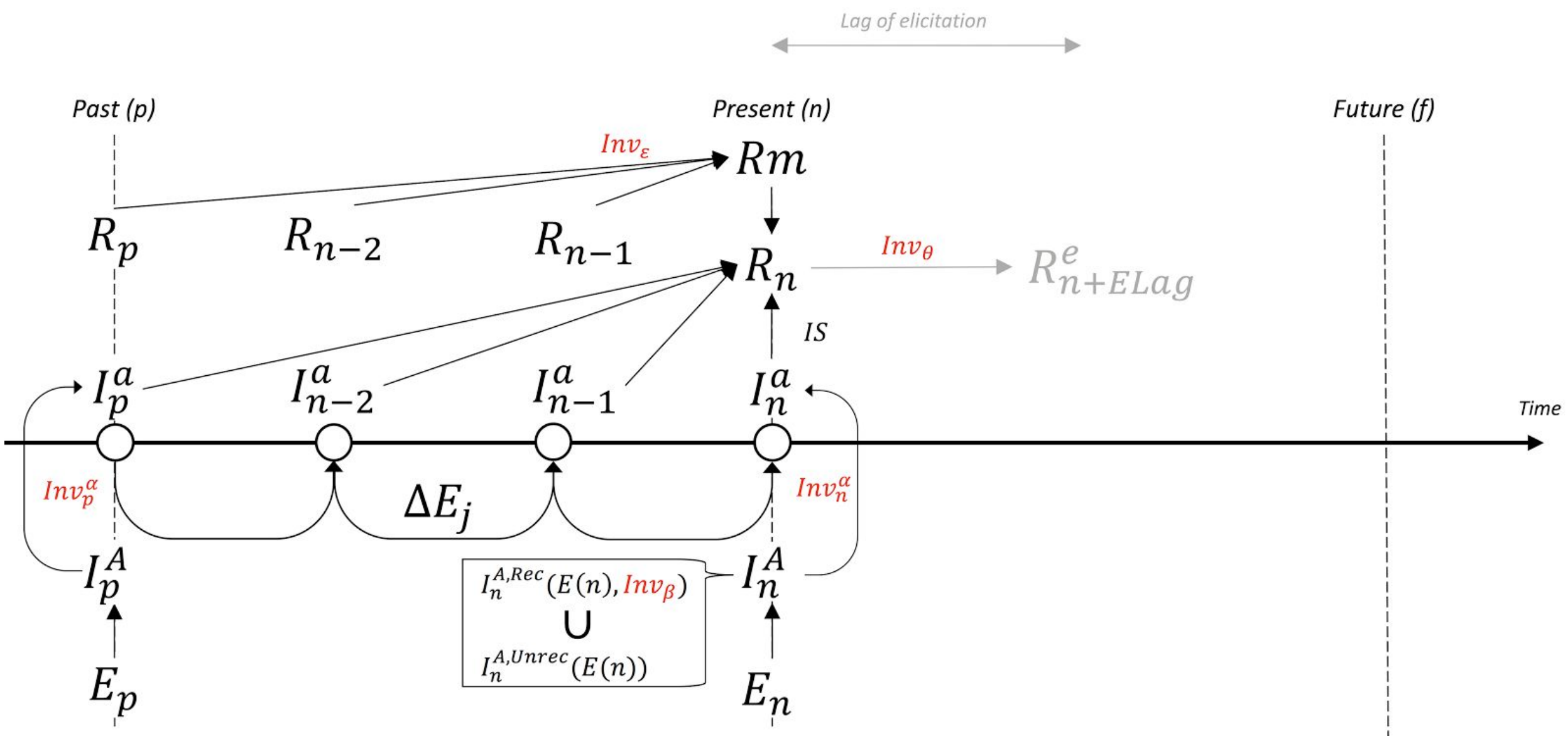

*Figure 6 - Elicited Requirements and Lag of Elicitation*

# 8. Freezing Requirements Casts a Chill

*The basic purpose of RE is to identify and analyse requirements in order to specify the behavior of a system*. Following Zave and Jackson seminal work on the requirements problem, engineers therefore expect to define $S$ such that $E, S \rightarrow R$ . To take into account the different propositions we made in this paper, the problem could be stated as follows: *to identify and analyse present requirements in order to specify the behavior of a future system.* In other words, define $S_f$ with $S_f = f(R^e_n, Inv_\mu)$ , where $Inv_\mu$ is the amount of resources invested in the analysis and specification activities. The introduction of a time parameter in the formulation of the requirements problem raises a series of questions. The most significant one is probably the following: “what are the implications of specifying a future system based on present information, in an environment that is constantly changing?”.

To answer this question, we need to have a closer look at the $S_f = f(R^e_n, Inv_\mu)$ formula. To manage the present specification of a future systems, it is common to freeze requirements; present requirements ($R_t$) are elicited ($R^e_t$) and then freezed to enable the definition of a specification $S_f$, assuming that these requirements will not evolve over time. The problem then becomes the definition of $S_f = f(R^e_t, Inv_\mu)$, assuming $R^e_t = R^e_f$, and as a consequence $E^a_t = E^a_f$. Remembering discussion from Section 1, we know that $E^a_f = E^a_t \cup (j : n \rightarrow f) \Delta E_j$ ,i.e., the future description of the environment equals the present description of the environment together with some changes that occurred between the present and the future. To assume $E^a_t = E^a_f$ therefore amounts to assume that $\cup (j : n \rightarrow f) \Delta E_j$ , i.e., no change occurs between the moment requirements are elicited and the moment the system is specified. Note that we call the difference between $n$ and $f$ the *Lag of Specification*.

The strategy above echoes many theories in decision making and non-monotonic reasoning suggesting that people confronted to uncertain decision settings resort to specific heuristics, such as the Default Logic. Simply stated, using default logic amounts to claim that, by default of information suggesting a change can occur, it can simply be assumed that no change will occur at all. In our case, it amounts to assume that, by default of elements indicating changes could occur, we simply assume no change will occur at all. With such point of view, requirements engineering amount to detect as much potential future changes, deal with them and adapt $R^e_n$ accordingly, and assume no other change will occur. The negative implication of such default reasoning strategy is obvious; there is a risk of assuming no change will occur while in fact a change will occur. This risk of an erroneous default assumption increases as the Lag of Specification gets bigger. This is depicted in Figure 7.

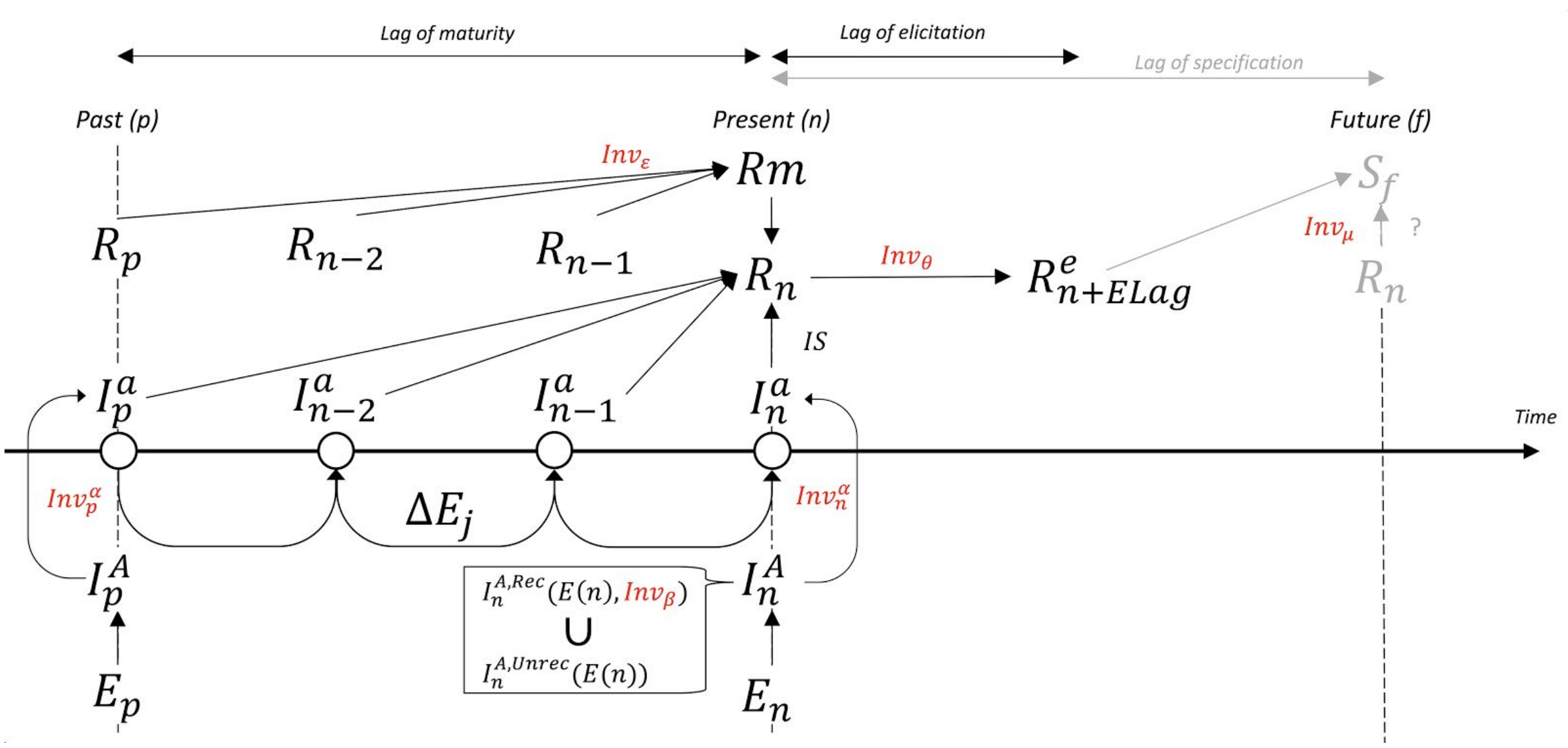

*Figure 7 - Future Specification and Lag of Specification*

# 9. First Steps Towards Economics of Requirements

Introducing a time dimension and the idea of requirements learning in the formulation of the requirements problem enabled to introduce a series of parameters which, until now, have not been studied by the RE community. The addition of these parameters raises the question of which value they should take in order to optimize the result of the RE process, how they should be managed and balanced. We do not provide an answer to this question, because it is a complex one, which we expect considerable future research on. However, we identify two significant trade-off that should be accounted for by requirements engineers.

# 9.1. Financial Trade-Off of Requirements

Throughout this paper, we introduced a series of investment parameters. Altogether, these parameters act as constraints on the way the requirements problem can be solved; in fact, the sum of all investments consented in a RE project should stay below the limit of budget. The goal is always to maximize the quality of the final specification of a system, and to do that, resources can be allocated to various efforts (they are all listed in Table 1). The requirements problem with a time dimension therefore becomes an optimisation problem:

$$MAX\ Quality(S_f)$$
$$such\ that\ Inv_\alpha + Inv_\beta + Inv_\varepsilon + Inv_\theta + Inv_\mu \leq Budget$$

| Investment | Actor | Description |
|---|---|---|
| α | Stakeholder | Resources allocated to access old states of the attainable |
| | | environment; searching for recorded information, focusing and concentrating to extract old $I^A$ from memory. |
| $\beta$ | Stakeholder | Resources allocated to the transcription of attainable information for the purpose of accessing it in the future, when needed (wiki page, knowledge base, memos, models). |
| $\varepsilon$ | Stakeholder/ Engineer | Resources allocated to the validation and testing of temptative requirements in order to gain in maturity |
| $\theta$ | Engineer | Resources allocated to the elicitation and documentation of requirements (focus is on elicitation, negotiation, validation) |
| $\mu$ | Engineer | Resources allocated to the treatment of requirements and to the specification of the system (focus is on analysis, modeling, specification). |

*Table 1 - The different types of Investment in a RE Project*

It is interesting to note that all costs described in our model are introduced because they reflect potential consumption of resource consented in each step of a classical RE process. Typically, RE starts with requirements elicitation ($\alpha$, $\theta$), proceeds with information transcription and documentation ($\beta$), pursue with the definition of models ($\mu$) and validation of requirement ($\varepsilon$). In each RE activity, costs can be either assigned to engineers, to stakeholders or to both of them. In practice, there may be additional costs, or it may be possible to split some of the costs above in more specific variables, but this is out of scope in this paper.

## 9.2. Temporal Trade-Offs of Requirements

A second tradeoff exists in requirement engineering in general, and which arises due to two antagonistic trends in the formulation of the requirement problem.

The first trend derives from the fact that requirements cannot be elicited instantaneously. This means that, anytime requirements must be collected from a stakeholder, the engineer needs some time to apply elicitation techniques, formalize and document those requirements. This amounts to say that in practice, it is not feasible to collect accurately $R_n$ at time $n$. At best, we can collect $R_n$ at time $n+LE$, where $LE$ is the Lag of Elicitation we discussed in Section 6. The shortest $LE$, the higher is the risk that $R_n^e$ (the result of the elicitation) mistakenly reflect actual stakeholders' requirements. On the contrary, higher values of $\Delta$ help in reducing this risk. Following this trend, and all else unchanged, we would try to increase the lag of elicitation and try to produce specifications of the system as late as possible, when all requirements are correctly elicited from stakeholders and documented by engineer.

The second trend derives from the fact that requirements engineers seem to make an extensive use of the default logic; specifications $S_f$ are produced based on $R_n$. This approach is valid only if there is no change in the environment $E$ between the moment requirements start to be learned by stakeholders and the moment specifications are produced by engineers. In other words, if the default assumption that the environment won't change is satisfied. The risk associated to this assumption is relatively small if we work on small Lags of Specification ($LS$). However, it likely increases as the Lag of Specification increases. Following this trend, and all else unchanged, we would decrease the Lag of Specification as much as possible and produce specifications as soon as the requirements start to be learned.

In practice, these two effects happen simultaneously; requirements are elicited while the environment is changing. The previous *ceteris paribus* reasoning therefore does not hold, and both trends have to be taken into account, simultaneously. Classical waterfall approaches opt for a long Lag of Elicitation and therefore a long Delay of Specification. This way, the risk of defects in elicitation is minimized, yet the risk of defects in specification is maximal. This has been recognized several time by research and practitioners, who came up with another strategy; reducing as much as feasible the Lag of Elicitation and therefore the Delay of Specification. This way, evolution in the environment is too small to bring defects in the specification. This is the strategy adopted by Agile methodologies. One recognized drawback of Agile approach, however, is that churn and rework are high, interactions with stakeholders are constant and extremely demanding and the short paced learning process is subject to many defects. The temporal trade-off arising from present discussion opens the door for a more nuance approach; maybe. While we have no cues for now about the answer to that question, we believe it worths more scientific investigation. The final formulation of the requirements problem with time become:

$$MAX\ Quality(S_f)$$
$$such\ that\ \ LS\,(LE) + LE \ \leq Delay$$

# 10. Conclusion

In this paper, we discussed the requirements problem and how it changes if we introduce a time component in it. We discuss the definition of the environment, requirements and specifications with a time component. The consequence of this new component is the introduction a series of investment variables which have to be taken into account when managing an RE project. The revised requirements problem also raise an important temporal trade-off, which has to be dealt with simultaneously with the budget trade-off.

# 11. References

- (B1979) Barry W. Boehm. 1979. Software engineering-as it is. In *Proceedings of the 4th international conference on Software engineering* (ICSE '79). IEEE Press, Piscataway, NJ, USA, 11-21
- (SG2016) Standish Group International. The Chaos Report 2016. http://www.standishgroup.com/outline
- (ZJ1997) Pamela Zave and Michael Jackson. 1997. Four dark corners of requirements engineering. *ACM Trans. Softw. Eng. Methodol.* 6, 1 (January 1997), 1-30. DOI=http://dx.doi.org/10.1145/237432.237434
- (EBJ2011) Ernst, Neil A., Alexander Borgida, and Ivan Jureta. "Finding incremental solutions for evolving requirements." *Requirements Engineering Conference (RE), 2011 19th IEEE International*. IEEE, 2011.
- (EBJM2012) Ernst, Neil, et al. "Agile requirements evolution via paraconsistent reasoning." *Advanced Information Systems Engineering*. Springer Berlin/Heidelberg, 2012.
- (JBEM2015) Jureta, Ivan J., et al. "The requirements problem for adaptive systems." *ACM Transactions on Management Information Systems (TMIS)* 5.3 (2015): 17.
- (J2017) Jureta, Ivan J. "What Happens to Intentional Concepts in Requirements Engineering If Intentional States Cannot Be Known?." *International Conference on Conceptual Modeling*. Springer, Cham, 2017.